\documentclass[12pt]{article}
\usepackage{a4}
\usepackage[cp1251]{inputenc}
\usepackage[english, ]{babel}
\usepackage[dvips]{graphicx}
\usepackage{graphicx}
\begin{document}
\title{The polarization tensor of neutral gluons\\
 in external fields at high temperature}
\author{V.V. Skalozub$^a$, A. V. Strelchenko$^b$\footnote{Permanent address: Dniepropetrovsk National University,
St. Naukova 13, Dniepropetrovsk  49050, Ukraine.}\\
$^a$Dniepropetrovsk National University\\
St. Naukova 13, Dniepropetrovsk  49050, Ukraine\\
$^b$University of Leipzig, Institute for Theoretical Physics\\
Augustusplatz 10/11, 04109 Leipzig, Germany\\ 
skalozub@ff.dsu.dp.ua, 
strelchenko@itp.uni-leipzig.de}
\date{}
\maketitle
\thispagestyle{empty}
\begin{abstract}
The one-loop polarization operator of neutral gluons in the background constant Abelian isotopic, $H_{3}$, and hypercharge, $H_{8}$, chromomagnetic fields combined with $A_0$ electrostatic potential at high temperature is calculated. The case when $A_0=0$ is investigated separately.
The proper time method is applied. It is found that neutral gluons do not acquire magnetic masses in the background fields, in contrast to the charged ones. The application of the results are discussed.
\end{abstract}
\section{Introduction}

Investigation of the deconfinement phase of QCD  remains of considerable interest for high-energy physics and cosmology. Among the most important objects here is a gluon polarization tensor (PT) containing information on the excitation spectrum of quark-gluon plasma. First the QCD PT was calculated and investigated in one-loop order of perturbation theory at $T\not=0$ by Kalashnikov and Klimov \cite{KK1}, \cite{KK2} (see also surveys \cite{Kal} and \cite{KSC}, \cite{Klsh} where the results on higher order contributions are discussed). As it has been shown, the  space components of the one-loop gluon propagator 
calculated within a standard perturbation  theory possesses a fictitious infrared pole at $k_4 = 0 , \bar{k} \sim g^2 T$ which could not be removed by any further resummations. These
 infrared divergencies of the termal Green functions provide the most challenging difficalties in understanding the internal structure of perturbative finite temperature  QCD.
It is believed, however, that formation of some condensate fields, such as a uniform "colour" magnetic field ($H_{c}=const.$) or electrostatic potential  (so-called $A_{0}$ condensate), can improve the infrared properties of the theory.  These condensate fields may arise in the deconfinement phase of QCD due to a peculiar dynamics of non-Abelian gauge fields, as it was argued by several authors \cite{Skal}-\cite{Sk}. In the paper by Kalashnikov \cite{Kalash94} it was demonstrated, in particular, that the $A_0$ condensate shifts the fictitious pole and introduces the gluon magntic mass of the order $m^2 \sim g^4 T^2$. At the same time,  in Ref.\cite{SkSt2} it was discovered  that in the presence of the external Abelian chromomagnetic fields  $H$ the transversal charged gluons acqure a magnetic mass $ m^2_{magn.} \sim g^2 ~\sqrt{gH} ~T $ which is generated within the one-loop polarization operator. It  acts to stabilize the external field.  In Ref.\cite{SkSt1} it was found  within the $SU(3)$ gluodynamics that at high temperature a specific combination of the Abelian hypercharge, $H_8$, and isotopic spin, $H_3$ , fields is generated and is stable  due to this magnetic mass. It is also of the order $ \sim g^4 T^2$.  The tachyonic (unstable) modes of the transversal charged gluons, which appear in the energy spectrum of the charged vector particles when the homogenious magnetic field is applied to the system, are  removed by these high-temperature radiative corrections. Moreover, an imagenary part of the
effective potantial (EP) of the background fields is cancelled if the contribution of the daisy diagrams with this magnetic mass is taken into consideration. Hence, one has to  believe that the non-trivial configuration  of the classical magnetic   fields $H_{3}$ and $H_{8}$ is generated in the deconfinement phase.

It is interesting  to see in actual calculations whether or not the magnetic mass of the neutral gluons is generated in the external field at high temperature. Actually,  this is not expected  because on general theoretical grounds the fields belonging to the Abelian projection of the non-Abelian groups remain massless. It is also important to know whether or not the fictitious pole of the neutral gluons is preserved when a magnetic field and $A_0$ is present in the system.

The aim of the present paper  is to calculate the one-loop polarization operator of the neutral gluons in SU(3) gluodynamics in the external fields $H_3$ and $H_8$   and $A_0$( or $A_4$) electrostatic potential at high temperature and
  check whether the full propagator of neutral gluons $Q_{3}$ and $Q_{8}$ contains the fictitious pole leading to the infrared instability. If this is not the case, one is able to conclude that the formation of the condensate fields play the role of an infrared regulator  and the transversal components of neutral gluons are unscreened. It is necessary to note that at zero temperature this problem was investigated in Ref. \cite{NO}.
 We will begin with the case when both the chromomagnetic fields and the electrostatic potentials are present in the system. Then the case of $gA_{0}=0$ will be separately analysed.
 We will restrict our consideration to the one-loop approximation.  To evaluate integrals over a three-dimantion momentum the Fock-Schwinger proper time method will be applied.
 The most essential steps of calculation are given in the Appendix 1.

\section{Calculation of the polarization operator}

We start our analysis
with the expression of the Lagrangian of neutral gluons in (Euclidean) $SU(3)$  gluodynamics:
\begin{eqnarray}\label{one}
&L_{neut.gl.}=-\frac{1}{4}Q_{\mu\nu}^{3}Q_{\mu\nu}^{3}
-\frac{1}{4}Q_{\mu\nu}^{8}Q_{\mu\nu}^{8}
-\frac{1}{2}(\partial_{\mu}Q_{\mu}^{3})(\partial_{\nu}Q_{\nu}^{3})
-\frac{1}{2}(\partial_{\mu}Q_{\mu}^{8})(\partial_{\nu}Q_{\nu}^{8})
&\nonumber\\
&~~~~~~~~+i  gQ_{\mu\nu}^{3} W_{1\mu}^{+}W_{1\nu}^{-}
+igQ_{\mu}^{3}(W_{1\nu}^{+}(\partial_{\mu}W_{1\nu}^{-}-\partial_{\nu}W_{1\mu}^{-})
-(h.c.)) &\nonumber\\
&~~~~~~~~~~~~+i \sqrt{\frac{3}{2}} g((
Q_{\mu\nu}^{8}+\frac{1}{\sqrt{6}}Q_{\mu\nu}^{3}) W_{2\mu}^{+}W_{2\nu}^{-}
+(
Q_{\mu\nu}^{8}-\frac{1}{\sqrt{6}}Q_{\mu\nu}^{3}) W_{3\mu}^{+}W_{3\nu}^{-})&\nonumber\\
&~~~~+i\sqrt{\frac{3}{2}}g(Q_{\mu}^{8}+\frac{1}{\sqrt{6}}Q_{\mu}^{3})
(W_{2\nu}^{+}(\partial_{\mu}W_{2\nu}^{-}-\partial_{\nu}W_{2\mu}^{-})
-(h.c.))
&\nonumber\\
&~~~~+i\sqrt{\frac{3}{2}}g(Q_{\mu}^{8}-\frac{1}{\sqrt{6}}Q_{\mu}^{3})
(W_{3\nu}^{+}(\partial_{\mu}W_{3\nu}^{-}-\partial_{\nu}W_{3\mu}^{-})
-(h.c.))&\nonumber\\
&~-g^{2}\Gamma_{\mu\nu\lambda\rho}Q_{\mu}^{3}Q_{\nu}^{3}(W_{1\lambda}^{+}W_{1\rho}^{-}
+\frac{1}{4}W_{2\lambda}^{+}W_{2\rho}^{-}
+\frac{1}{4}W_{3\lambda}^{+}W_{3\rho}^{-})
&\nonumber\\
&-\frac{3}{2}g^{2}\Gamma_{\mu\nu\lambda\rho}Q_{\mu}^{8}Q_{\nu}^{8}
(W_{2\lambda}^{+}W_{2\rho}^{-}+W_{3\lambda}^{+}W_{3\rho}^{-})
+L_{gh}.~~~~~~~&
\end{eqnarray}
Here  the following basis of charged gluons $Q_{\mu}^{a}$ ($a=1,2,4,5,6,7$)
\begin{equation}\label{two}
W_{1\mu}^{\pm}=\frac{1}{\sqrt{2}}(Q_{\mu}^{1}\pm iQ_{\mu}^{2}),
~ W_{2\mu}^{\pm}=\frac{1}{\sqrt{2}}(Q_{\mu}^{4}\pm iQ_{\mu}^{5}),
~ W_{3\mu}^{\pm}=\frac{1}{\sqrt{2}}(Q_{\mu}^{6}\pm iQ_{\mu}^{7}).
\end{equation}
is introduced.
The external potential is chosen in the form
$B_{\mu}^{a}=\delta^{a3}B_{3\mu}+\delta^{a8}B_{8\mu}$,
where\\
$B_{3\mu}=H_{3}\delta_{\mu2}x_{1}+\delta_{\mu4}gA_{3}$
and
$B_{8\mu}=H_{8}\delta_{\mu2}x_{1}+\delta_{\mu4}gA_{8}.$
In these formalae the notations $A_{3}$ and $A_{8}$ correspond accordingly to $A^{a=3}_{0}$ and $A^{a=8}_{0}$ electrostatic potentials.
The constant chromomagnetic fields are chosen  to be directed along the
third axis of the Euclidean space and  $a=3$ and $a=8$ of the colour
$SU_{c}(3)$-space: $F_{\mu\nu}^{a~ext}=\delta^{a3}F^{a=3}_{\mu\nu}+\delta^{a8}F^{a=8}_{\mu\nu}$,
$F^{a}_{12}=-F^{a}_{21}=H_{a}$,~$a=3,8$.
From the Lagrangian (\ref{one}) one can easely derive the diagrams describing 
 propagation of the
neutral gluons in the background fields.

In the one-loop approximation the PO of 
neural gluons is determined by the standard set of diagrams
in Fig. 1, where double wavy lines represent the Green function
$G_{r~\mu\nu} (x,y)$ for the charged gluons, dashed double lines
represent the Green function $D(x,y)$ for the charged ghost
fields. Thin wavy line corresponds to the neutral
gluon fields $Q_{\mu}^{3,8}$. In the operator form the above Green functions are
 given by the expressions (in Feynman's gauge)

\[G_{r=1~\mu\nu}(P)=-[P^2+2igF_{3\mu\nu}]^{-1},\]
\[G_{r=2,3~\mu\nu}(P)=-[P^2+\sqrt{6}i\lambda_{\pm} gF_{8\mu\nu}]^{-1},\qquad
D(P)=-\frac{1}{P^2},\]
$$\lambda_{\pm}=1\pm\frac{1}{\sqrt{6}}\frac{H_{3}}{H_{8}}.$$
To calculate the PO we make use of the proper time
representation and the Schwinger operator formalism \cite{Sch}.
The PO of the neutral gluons in the background fields at $T\not=0$ can be written as
\begin{equation}\label{three}
\Pi_{\mu\nu}^{a=3}=-g^2 T\sum_{P_4}
\int\frac{d^3P}{(2\pi)^3} (\Pi_{\mu\nu}(k,P)+\frac{1}{4}\widetilde{\Pi}_{\mu\nu}(k,P)),
\end{equation}
\begin{equation}\label{four}
\Pi_{\mu\nu}^{a=8}=-\frac{3}{2}g^2 T\sum_{P_4}
\int\frac{d^3P}{(2\pi)^3} \widetilde{\Pi}_{\mu\nu}(k,P),~~~~~~~~~~~~
\end{equation}
where
\begin{eqnarray*}
\Pi_{\mu\nu}(k,P)&=&\Pi^{r=1}_{\mu\nu}(k,P),\\
\widetilde{\Pi}_{\mu\nu}(k,P)&=&\sum_{r=2,3}\Pi^{r}_{\mu\nu}(k,P),\\
\Pi^{r}_{\mu\nu}(k,P)&=&\Bigl\{\Gamma_{r~\mu\alpha,\beta}(P,k)~
G_{r~\beta\lambda}(P)~\Gamma_{r~\nu\sigma,\lambda}(P,k)~G_{r~\sigma\alpha}(P-k)-2\delta_{\mu\nu}G_{r~\alpha\alpha}(P)+ \\
& &-2 \Bigl[(2P-k)_{\mu}D(P)(2P-k)_{\nu}D(P-k)-2
\delta_{\mu\nu}D(P)\Bigr]\Bigr\},\\
\Gamma_{\mu\alpha,\beta}&=&(2P-k)_{\mu} \delta_{\alpha\beta}-2(k_{\alpha}\delta_{\beta\mu}-k_{\beta}\delta_{\alpha\mu})
,
\end{eqnarray*}
 $P_{4} =2 \pi l T+gA_{3}$,
$P_{i}= i \partial_{i} +g B_{3i}$ for $r=1$
and
$P_{4} =2 \pi l T
+\sqrt{\frac{3}{2}}\mu_{\pm}gA_{8}$,\\
 $P_{i}= i \partial_{i} +
\sqrt{\frac{3}{2}}\lambda_{\pm} gB_{8i}$
for $r=2,3$,
respectively;
$l=0, \pm1,
\pm2, \ldots$, and
$$\mu_{\pm}=1\pm\frac{1}{\sqrt{6}}\frac{A_{3}}{A_{8}}.$$
 Assume now that the values of potantials
$A_{3}$ and $A_{8}$
satisfy the following conditions:
$gA_{3} \ll T$
and
$gA_{8} \ll T.$
 This is natural because the quantities
$gA_{3,8}$ are expected to be of order $g^{2}T$, as it is pointed out in Refs. \cite{ZSV},\cite{EZV} for $SU(2)$ case.
To  investigate the high temperature limit of (\ref{three}) and (\ref{four}) one can take the
$l=0$ term only  in the sum over $P_{4}$ \cite{Kal}.

To evaluate the expression for the PO let us apply the Schwinger
proper-time method modified for the case of high temperature. From technical point of veiw,
this case is similar to the
zero temperature one, so one may consult for more details, for example, to Refs.\cite{TE}- \cite{KhL}, where
the PO of photon as well as neutral gluon in the external (chromo)magnetic field were calculated at $T=0$ .
The basic steps of the calculating procedure are noted in the Appendix.
For simplisity it is convenient to introduce the following notations:
\begin{eqnarray}\label{five}
H_{\pm}=\sqrt{\frac{3}{2}}\lambda_{\pm}H_{8};~~
A_{\pm}=\sqrt{\frac{3}{2}}\mu_{\pm}A_{8};~~
m=\frac{(gA_{3})^{2}}{gH_{3}}, ~~m_{\pm}=\frac{(gA_{\pm})^{2}}{gH_{\pm}}.
\end{eqnarray}
Then the final result of evaluation (\ref{three}) and (\ref{four}) reads:
\begin{equation}\label{six}
\Pi_{ij}^{a=3,8}=(\delta_{ij}-\frac{k_{i}k_{j}}{\bar{k}^{2}})\bar{k}^{2}\Pi^{(1)}_{a=3,8}+
(B\bar{k})_{i}(B\bar{k})_{j}\Pi^{(2)}_{a=3,8},~~~~~~~~~~~~
\end{equation}
\begin{eqnarray}\label{seven}
\Pi_{i4}^{a=3,8}=-\Pi_{4i}^{a=3,8}
=i(B\bar{k})_{i}\Pi^{(3)}_{a=3,8},~~~~
\Pi_{44}^{a=3,8}=
 \Pi^{(4)}_{a=3,8}-\Psi_{a=3,8}.
\end{eqnarray}
Here the quantities $\Pi^{(i)}_{a=3,8}$, $i=1,..,4$, and $\Psi_{a=3,8}$ are
\begin{eqnarray}\label{eight}
\Pi^{(i)}_{a=3}=\Pi^{(i)}+\frac{1}{6}\Pi^{(i)}_{a=8},~~~
\Psi_{a=3}=\Psi+\frac{1}{6}\Psi_{a=8},
\end{eqnarray}
\begin{eqnarray}\label{nine}
\Pi^{(i)}=-\frac{g^{2}}{8 \pi^{3/2}}\frac{T}{\sqrt{gH_3}}
\int\limits_0^1 d u
\int\limits_0^\infty \frac{d x}{sh(x)} \sqrt{x}~~ exp[-\Phi-x m]
f^{(i)}(x,u),
\end{eqnarray}
\begin{eqnarray}\label{ten}
\Pi^{(i)}_{a=8}=-\frac{3 g^{2}T}{16 \pi^{3/2}}
\int\limits_0^1 d u
\int\limits_0^\infty \frac{d x}{sh(x)} \sqrt{x}
\Bigl\{\frac{l^{(i)}_{+}(x,u)}{\sqrt{gH_{+}}}e^{-\Phi_{+}-xm_{+}}
+\frac{l^{(i)}_{-}(x,u)}{\sqrt{gH_{-}}}e^{-\Phi_{-}-xm_{-}}\Bigr\},
\end{eqnarray}
\begin{eqnarray}\label{eleven}
\Psi=
 \frac{g^{2}}{4\pi^{3/2}}  \sqrt{gH_3}T
\int_0^\infty \frac{d x}{\sqrt{x}}
 \Bigl[\frac{2}{sh(x)}+4sh(x)\Bigr]e^{-xm},
\end{eqnarray}
\begin{eqnarray}\label{twelv}
\Psi_{a=8}=
\frac{3g^{2}}{8\pi^{3/2}}T
\int_0^\infty \frac{d x}{\sqrt{x}}
 \Bigl[\frac{2}{sh(x)}+4sh(x)\Bigr]\Bigl\{\sqrt{gH_{+}}e^{-xm_{+}}
+\sqrt{gH_{-}}e^{-xm_{-}}\Bigr\},
\end{eqnarray}
where
$$\Phi=xu(1-u)\frac{k_{3}^{2}}{gH_3}+k_{\perp}^{2}\frac{\zeta}{2gH_3},~~
\Phi_{\pm}=xu(1-u)\frac{k_{3}^{2}}{gH_{\pm}}+k_{\perp}^{2}\frac{\zeta}{2gH_{\pm}}$$
and
$$\zeta=\frac{ch(x)-ch(x(1-2u))}{sh(x)},~~k_{\perp}^{2}=k_{1}^{2}+k_{2}^{2}$$
Exact expressions for the functions $f^{(i)}$ and $l^{(i)}_{\pm}$ are adduced in the Appendix 2.
The matrix $B_{ij}$ is a usual two dimension antisymmetric tensor,
$$B_{ij}=  \epsilon _{ij} = \delta_{i2}\delta_{1j}-\delta_{i1}\delta_{2j}.$$
The spatial part of the PO is transversal manifestly, as it is required by gauge invariance. Note that $\Pi^{(i=3,4)}=0$ for $A_3=A_{\pm}=0$.

Now let us consider the high-temperature expansion,
$gH_{3,8}\ll T^{2}$, $(gA_{3,8})^{2} \ll T^{2}$,
 of the expressions in Eqs.(\ref{eight})-(\ref{twelv}).
Assuming that the quantities $gH_{3,8}$ and $(gA_{3,8})^{2}$ are of the same order of magnitude,
we investigate the two separate regimes: $\mid\bar{k}\mid \ll g^{2}T$ and $\mid\bar{k}\mid \ge gT$.
 In the former case, with the additional condition $k_{\perp}^{2} \ll gH_{3,8}$ and
$k_{3}^{2} < gH_{3,8}$, the main contributions to integrals come from the integration domain where
$x \gg 1$.
Carrying out integrations we obtain
\begin{eqnarray}\label{thirteen}
\Pi^{(i)}_{a=3}=\Pi^{(i)}(gH_3;\nu;m)+\frac{1}{6}\Pi^{(i)}_{a=8},
\end{eqnarray}
\begin{eqnarray}\label{fourteen}
\Pi^{(i=1,2)}_{a=8}=
\frac{3}{2}\Bigl\{\Pi^{(i=1,2)}(gH_{+};\nu_{+};m_{+})+
\Pi^{(i=1,2)}(gH_{-};\nu_{-};m_{-})
\Bigr\},
\end{eqnarray}
\begin{eqnarray}\label{fifteen}
\Pi^{(3)}_{a=8}=
\frac{3}{2}\Bigl\{\frac{A_{+}}{A_3}\Pi^{(3)}(gH_{+};\nu_{+};m_{+})+
\frac{A_{-}}{A_3}\Pi^{(3)}(gH_{-};\nu_{-};m_{-})
\Bigr\},
\end{eqnarray}
\begin{eqnarray}\label{sixteen}
\Pi^{(4)}_{a=8}=
\frac{3}{2}\Bigl\{(\frac{A_{+}}{A_3})^{2}\Pi^{(4)}(gH_{+};\nu_{+};m_{+})+(\frac{A_{-}}{A_3})^{2}
\Pi^{(4)}(gH_{-};\nu_{-};m_{-})
\Bigr\}.
\end{eqnarray}
Here $\nu=\frac{k_{3}^{2}}{4gH_3}$, $\nu_{\pm}=\frac{k_{3}^{2}}{4gH_{\pm}}$ and functions $\Pi^{(i)}(\alpha;\beta;\gamma)$ are represented by the following expressions:
\begin{eqnarray*}
\Pi^{(1)}(\alpha;\beta;\gamma)=-\frac{g^{2}T}{2\pi\sqrt{\alpha}}
\frac{1}{4(\gamma \beta+1)+\beta^2}[\frac{\frac{3}{2}-\gamma+\beta}{\sqrt{\gamma-1}}+\frac{\beta+\gamma-1}{\sqrt{\gamma+\beta}}],
\end{eqnarray*}
\begin{eqnarray*}
\Pi^{(2)}(\alpha;\beta;\gamma)=-\frac{g^{2}T}{8\pi\sqrt{\alpha}}
[\frac{2}{(\gamma-1)(\gamma+\beta-1)}-\\
\frac{1}{1+\beta(\gamma+\beta+1)}(\frac{1+\beta}{\sqrt{\gamma-1}}-\frac{1}{\sqrt{\gamma+\beta+1}})]-\Pi^{(1)},
\end{eqnarray*}
\begin{eqnarray*}
\Pi^{(3)}(\alpha;\beta;\gamma)=-\frac{g^{2}T}{\pi\sqrt{\alpha}}
\frac{g A_3 \sqrt{\gamma-1}}{[4(\gamma-1)+\beta]},
\end{eqnarray*}
\begin{eqnarray*}
\Pi^{(4)}(\alpha;\beta \rightarrow 0;\gamma)=-\frac{g^{2}T}{2\pi\sqrt{\alpha}}
\frac{(g A_3)^{2}}{(\gamma-1)^{3/2}},
\end{eqnarray*}
where according to (\ref{thirteen})-(\ref{sixteen}) instead of variables $\alpha$, $\beta$ and $\gamma$ one has to substitute $gH_3,~\nu,~m$ or $gHh_{\pm},~\nu_{\pm},~m_{\pm}$, respectively.
For $\Psi_{a=3}$ and $\Psi_{a=8}$ we have:
\begin{eqnarray}
\Psi_{a=3}=\Psi(gH_3;m)+\frac{1}{6}\Psi_{a=8},
\end{eqnarray}
\begin{eqnarray}\label{eighteen}
\Psi_{a=8}=\frac{3}{2}(\Psi(gH_{+};m_{+})+\Psi(gH_{-};m_{-})),
\end{eqnarray}
where
\begin{eqnarray*}
\Psi(\alpha;\gamma)=\frac{g^2 T}{\pi}\sqrt{\alpha}[\frac{1}{\sqrt{\gamma+1}}+\sqrt{\gamma-1}].
\end{eqnarray*}
 For the values $m=1$ and/or $m_{\pm}=1$ the functions $\Pi^{(i=1,2)}$ and $\Pi^{(4)}$ become divergent whereas $\Pi^{(3)}$ is equal to zero.

In the case of $\mid\bar{k}\mid \ge gT$ and $k_{\perp}^2 \gg (gA_{3,8})^2$ (but $m>1$ and $m_{\pm}>1$), the main contributions to integrals come from the  region $x\sim 0$. Expanding the integrand
 functions into the power  series over the veriable $x$, one can obtain for the spatial components  (\ref{six}):
\begin{equation}\label{nineteen}
\Pi_{ij}^{a=3,8}\sim (\delta_{ij}-\frac{k_{i}k_{j}}{\bar{k}^{2}})\bar{k}^{2}
\Pi^{(1)}_{a=3,8}+
(B\bar{k})_{i}(B\bar{k})_{j}\Pi^{(2)}_{a=3,8}.~~~~~~~~~~~~
\end{equation}
Here
\begin{equation}\label{twenty}
\Pi^{(1)}_{a=3}=-\frac{21}{16}C,~~ \Pi^{(2)}_{a=3}=
\frac{3}{4}C,~~\Pi^{(1)}_{a=8}=-\frac{21}{8}C,~~\Pi^{(2)}_{a=8}=
\frac{3}{2}C,~~~C=\frac{g^2 T}{k_{\perp}}.
\end{equation}
It is remarkable that the quantities (\ref{twenty}), which are, of cause, only the leading terms of perturbative expansion,  do not depend upon the condensate fields. For the momentum scale $k_{\perp} \sim T$ the constant $C$ is of order $g^2$ and, therefore, perturbative theory is actually governed by the parameter $g^2$. However, for the scale $k_{\perp} \sim gT \ll T$ the effective expantion parmeter becomes  $g$. Hence one can see that perturbative features of the model are aggravated with the decreasing $k_{\perp}$.

\section{Polarization operator in the extarnal magnetic fields}

In this section we  consider the PO  in the external chromomagnetic fields $H_{3,8}$ (but $gA_{3,8}=0$). We merely put the parameters $A_3,~A_{\pm}$, in the Eqs.
(\ref{nine})-{\ref{twelv}) equal to zero. In this case the integrands in the r.h.s. of Eqs.
(\ref{nine})-(\ref{twelv})  are nonanalytical for large $x$. 
To ensure the convergence of integrals with respect to $x$ one has to rotate the integration
 contour by the standard rule: $x \rightarrow ix$. Then, assuming again that  $k_{\perp}^{2} \ll gH_{3,8}$ and
$k_{3}^{2} \ll gH_{3,8}$, the main contributions  come from large $x$ and we obtain:
\begin{eqnarray}\label{twentyone}
\Pi^{(i)}_{a=3}=\Pi^{(i)}(gH_3;\nu)+\frac{1}{6}\Pi^{(i)}_{a=8},
\end{eqnarray}
\begin{eqnarray}\label{twentytwo}
\Pi^{(i=1,2)}_{a=8}=
\frac{3}{2}\Bigl\{\Pi^{(i=1,2)}(gH_{+};\nu_{+})+
\Pi^{(i=1,2)}(gH_{-};\nu_{-})
\Bigr\},
\end{eqnarray}
where
\begin{eqnarray*}
\Pi^{(1)}(\alpha;\beta)=-\frac{g^{2}T}{2\pi\sqrt{\alpha}
(1+4\beta^2)}[3\beta^2-\frac{1}{2}+ i(3\beta^2+\frac{1}{2})],
\end{eqnarray*}
\begin{eqnarray*}
\Pi^{(2)}(\alpha;\beta)=-\frac{g^{2}T}{2\pi\sqrt{\alpha}}
[\frac{1}{4\sqrt{\beta+1}(1+\beta(\beta+1))}-\frac{2(1-2\beta)}{1+4\beta}+\\
i (\frac{1}{2(1-\beta)}+\frac{1+2\beta}{1+4\beta}
-\frac{1+\beta}{4(1+\beta(\beta+1))})]-\Pi^{(1)}.
\end{eqnarray*}
The Debye masses of neutral gluons are
\begin{eqnarray}\label{twethree}
Re(\Pi_{44}^{a=3})=
\frac{g^{2}}{\pi}T\Bigl[\sqrt{gH_{3}}+\sqrt{gH_{8}}\frac{1}{4}(\sqrt{\lambda_{+}}+\sqrt{\lambda_{-}})\Bigr]
\end{eqnarray}
and
\begin{eqnarray}\label{twentyfour}
Re(\Pi_{44}^{a=8})=
\frac{3g^{2}}{2\pi}T\sqrt{gH_{8}}(\sqrt{\lambda_{+}}+\sqrt{\lambda_{-}}).
\end{eqnarray}
Quantities (\ref{twentyone}) and (\ref{twentytwo}), as well as $\Pi_{44}$, include imaginary parts reflecting the existence of the  tachyonic mode in the tree-level spectrum of charged gluons. 
   It should be noted, that expressions for $\Pi_{44}$ represent the next-to-leading terms.
 To calculate the leading terms one has to perform  summation over the discrete frequencies $P_{4}$.

\section{Discussion}
 To discuss the results obtained, let us consider the full propagator of the neutral gluons $Q_{\mu}^{a=3,8}$. To one-loop
order the transversal part of the propagator spatial components  has the following structure
 \begin{eqnarray}\label{twentyfive}
G_{ij}^{tr}=(\delta_{ij}-\frac{k_{i}k_{j}}{\bar{k}^{2}})\frac{1}{\bar{k}^{2}(1+\Pi^{(1)})}-~~~~~~~~~~~~~~~~~~~~~~~~~~~~~~~~~~~~~~~~\nonumber\\
-\frac{(B\bar{k})_{i}(B\bar{k})_{j}}{\bar{k}^{2}}\frac{\Pi^{(2)}}
{[\bar{k}^{2}(1+\Pi^{(1)})+k_{\perp}^{2}\Pi^{(2)}](1+\Pi^{(1)})},
\end{eqnarray}
where the functions $\Pi^{(1,2)}$ are given by Eqs.(\ref{thirteen}), (\ref{fourteen}), (\ref{twenty}), (\ref{twentyone}) and (\ref{twentytwo}). In the case of $gA_{3,8}\not=0$ (see Eqs. (\ref{thirteen}), (\ref{fourteen}) and (\ref{twenty}) ), the full propagator of does not contain a non-trivial pole.
Hence, one has to conclude that the neutral gluons do not acquire magnetic masses in the presence of the background fields $A_{3,8}$ and $H_{3,8}$.

Here a more serious problem arises.  Namely, if the condensate fields are of the order $g^2 T$, as it was argured in Refs. \cite{Skal}-\cite{EZV}, then, for the case of $k_{\perp}^{2} \ll \sqrt{gH_{3,8}}$ (see (\ref{thirteen})-(\ref{sixteen})), the factors $\frac{g^2 T}{\sqrt{gH_{3,8}}}$ appearing in (\ref{thirteen})-(\ref{sixteen}) turn out to be of order $O(1)$ and the perturbative expansion breaks down for the momentum scale $k_{\perp} \ll g^2 T$.  Therefore one cannot explore the infrared region ($\bar{k} \rightarrow 0$) by usual perturbative methods and our conclusion is  valid for the scale $k_{\perp} \ge g T$, only. In this region perturbation theory is reliable (see Eqs. (\ref{nineteen})-(\ref{twenty}) and the text below).

In the case of the chromomagnetic fields been taken into consideration the quantities $\Pi^{(1,2)}$ were found to be complex, and Eq. (\ref{twentyfive}) can be rewritten as
\begin{eqnarray}\label{twentysix}
G_{ij}^{tr}=(\delta_{ij}-\frac{k_{i}k_{j}}{\bar{k}^{2}})\frac{1+Re\Pi^{(1)}-i~Im\Pi^{(1)}}{\bar{k}^{2}
[(1+Re\Pi^{(1)})^2+(Im\Pi^{(1)})^2]}-~~~~~~~~~~~~~~~~~~\nonumber\\
-(B\bar{k})_{i}(B\bar{k})_{j}
\frac{(1+Re\Pi^{(1)}-i~Im\Pi^{(1)})\Pi^{(2)}}{\bar{k}^{2}
[(1+Re\Pi^{(1)})^2+(Im\Pi^{(1)})^2]}
\times~~\nonumber\\
\times\frac{(\bar{k}^{2}(1+Re\Pi^{(1)})+k_{\perp}^{2}Re\Pi^{(2)}-i~(\bar{k}^{2}Im\Pi^{(1)}+k_{\perp}^{2}Im\Pi^{(2)}))}
{[(\bar{k}^{2}(1+Re\Pi^{(1)})+k_{\perp}^{2}\Pi^{(2)})^2+(\bar{k}^{2}Im\Pi^{(1)}+k_{\perp}^{2}Im\Pi^{(2)})^2]}.
\end{eqnarray}
This expression has also a pole  at $\bar{k}^{2}=0$, only. However, the imaginary part that arises in  Eq. (\ref{twentysix}) has a "tachyonic" origin, as it was mentioned above.
Really, the calculation of  $\Pi_{ij}$ (as well as $\Pi_{44}$) has been carried out with the bare propagators of the charged gluons substituted into internal lines of  diagrams. This results in a non-analyticity of integrands with respect to the variable $x$ in the $\Pi_{\mu\nu}$. In this sence the carried out one-loop calculation of the PO appears to be insufficient: to obtain a correct independent of the imaginary part expressions for (\ref{twentysix}), the  charged gluon propagators accounting for the magnetic mass derived in the paper \cite{SkSt1} must be used. But now, when we know the origin of the imaginary part, it does not matter when the problem on the magnetic mass of the neutral gluons is investigated.

In the present paper it was straightforwardly demonstrated  that the transversal neutral gluon fields are not screened by thermal fluctuations if the non-trivial condensates  present in the QCD deconfinement phase.
 We arrived at the following picture when the assumed formation of condensate fields $gA_{3,8}$ and $gH_{3,8}$ determine the effective masses of the charged gluons while the neutral spatial components do not acquire magnetic masses in the fields. It is resonable to suppose that this picture will be also valid when only chromomagnetic fields $gH_{3,8}$ are generated in the system although higher-order contributions to the neutral gluon PT must be taken into account in this case.
It is worth to emphasize that in the infrared region, $k \rightarrow 0$, the full propagator (\ref{twentyfive}) does not contain the "fictitious" pole. This is in contrast to the case of trivial vacuum \cite{KK1}, \cite{Kal}.  
\section{Acknowledgments}
One of us (A.S.) thanks M. Bordag for helpfull discussions and the graduate college Quantenfieldtheorie at the University of Leipzig for support and friendly enviroment. This work (V.S.) was supported in part by the grant from DFG No: UKR 427/17/03.

\newpage

\section{Appendix 1}

To illustrate the basic stages of evaluating the PO (\ref{three}-\ref{four}) let us consider the integral:
\begin{eqnarray*}
I_{ij}=\frac{g^2}{\beta}
\int\frac{d^3P}{(2\pi)^3} \Pi_{ij}(\bar{k},\bar{P}),
\end{eqnarray*}
which represents the contribution of the charged fields $W^{\pm}_{r=1}$ (and the corresponding ghosts) to the $\Pi_{ij}^{a=3}$  at high temperature. The rest components of the tensor  $I_{\mu\nu}$ are calculated analogously.
Following the standard procedure we introduce a proper time for
each propogator appearing in $\Pi_{ij} (\bar{k}, \bar{P})$:
 $$D(\bar{P})=-\int_0^\infty d s  e^{-s \bar{P}^{2}},$$
 $$G_{r=1~\mu\nu}(\bar{P})=-\int_0^\infty d s  e^{-s\bar{P}^2-2igF_{3\mu\nu}s},$$
  Then, the whole expression for $I$ can be rewritten in the form:
\begin{eqnarray*}
I_{ij}=\frac{g^2}{\beta}\int_0^\infty d s_{1}d s_{2}
\int\frac{d^3P}{(2\pi)^3}
exp[-(s_{1}+s_{2})(gA_{3})^{2}]~~~~~~~~~~~~~~~~~~~~~~~~~~~~~~~~~~~~~~~~~~\nonumber\\
\Bigl\{\Bigl[\Gamma_{1~il,m}(\bar{P},\bar{k})~
\Lambda_{mn}(\sigma_{1})~\Gamma_{1~js,n}(\bar{P}',\bar{k})~\Lambda_{sl}(\sigma_{1}')-2(2\bar{P}-\bar{k})_{i}(2\bar{P}'-\bar{k})_{j}\Bigr]
\theta_{r=1}\Bigr\}
~~\nonumber\\
-2\frac{g^2}{\beta}\delta_{ij}\int_0^\infty d s \int\frac{d^3P}{(2\pi)^3}
 \Bigl[Tr\Lambda(\sigma_{1}'')-2\Bigr]e^{-s(\bar{P}^2+(gA_{3})^{2})},
\end{eqnarray*}
where
$\Gamma_{1~il,m}=(2P-k)_{i} \delta_{lm}-2(k_{l}\delta_{mi}-k_{m}\delta_{li})$ is the vertex factor,
\begin{eqnarray*}
\theta_{r=1}=e^{-s_{1}\bar{P}^2}e^{-s_{2}(\bar{P}-\bar{k})^2},~~~
\Lambda_{ij}(x)=R_{ij}-B_{ij}^{2}ch(x)-iB_{ij}sh(x)
\end{eqnarray*}
and variables $\sigma_{1}, ~\sigma_{1}',~ \sigma_{1}''$ are
$$\sigma_{r=1}=2igH_{3}s_{1},~~~
\sigma_{r=1}'=2igH_{3}s_{2},~~~
\sigma_{r=1}''=2igH_{3}s.$$
We introduced  the following designation: $P_{i}'=(exp[-2igF_{3}s_{1}]P)_{i}$,
$P_{i}= i \partial_{i} +g B_{3i}$.
The matrixes $R_{ij}$, $B_{ij}$ and $B_{ij}^{2}$ are:
$$R_{ij}=\delta_{i3}\delta_{3j},~~~ B_{ij}=\delta_{i2}\delta_{1j}-\delta_{i1}\delta_{2j},~~~ B_{ij}^{2}=B_{il}B_{lj}.$$
Next,
three-dimensional integration with respect to $\bar{P}$ in
$I_{ij}$ is carried out by means of the transition to the
conjugate variable  $X_i^\prime$: $$[X_i,P_j]=i\delta_{ij}.$$ By
using the eigenstates of the operator $X_i$ as determined by the
condition $X_i^\prime=0$, the integral over $\bar{P}$ can be
represented as
 $$\int\frac{d^3P}{(2 \pi)^3}f(\bar{P})=\langle
\bar{X}^\prime=0\mid f(\bar{P})\mid \bar{X}^\prime=0\rangle .$$
Hence, performing the following transformation of variables $s_1$ and $s_2$:
 $s_{1}=s(1-u)$, $s_{2}=su$, we have for $I_{ij}$ :

\begin{eqnarray*}
I_{ij}=\frac{g^{2}}{\beta}
\int\limits_0^1 d u
\int\limits_0^\infty d s s~~ exp[-s(gA_3)^2]
~~~~~~~~~~~~~~~~~~~~~~~~~~~~~~~~~~~~~~~~~~~~~~~~\nonumber\\
\langle
\Bigl[\Gamma_{1~il,m}(\bar{P},\bar{k})~
\Lambda_{mn}(\sigma_{1})~\Gamma_{1~js,n}(\bar{P}',\bar{k})~\Lambda_{sl}(\sigma_{1}')-2(2\bar{P}-\bar{k})_{i}(2\bar{P}'-\bar{k})_{j}\Bigr]
\theta_{r=1}\rangle
~~\nonumber\\
-2\frac{g^2}{\beta}\delta_{ij}\int_0^\infty d s
 \Bigl[Tr\Lambda(\sigma_{1}'')-2\Bigr]exp[-s(gA_3)^2]\langle e^{-s\bar{P}^2}\rangle,
\end{eqnarray*}
For convenience we use a notation $\langle
\bar{X}^\prime=0\mid ...\mid \bar{X}^\prime=0\rangle=\langle...\rangle.$
Now one need to calculate the quantities $\langle \theta_{r=1}\rangle$, $\langle P_{i}\theta_{r=1}\rangle$ and $\langle P_{i}P_{j}\theta_{r=1}\rangle$
according to the procedure described in Ref.\cite{Sch}. The result reads
 $$\langle P_{i}\theta_{r=1}\rangle=
\left(\frac{A}{D}\bar{k}\right)_i \langle \theta_{r=1}\rangle,$$
$$\langle P_{i}P_{j}\theta_{r=1}\rangle=
\left[\left(\frac{A}{D}\bar{k}\right)_i
\left(\frac{A}{D}\bar{k}\right)_j
-ig \left(\frac{F}{D^T}\right)_{ij}\right]\langle \theta_{r=1}\rangle,$$
$$\langle \theta_{r=1}\rangle =\frac{1}{(4\pi s)^{3/2}}
\frac{gH_{3}s}{sh(gH_{3}s)}
e^{-\Phi_3},$$
$$\Phi_3=k_{3}^{2}su(1-u)+k_{\perp}^{2}\frac{\zeta}{2gH_{3}},$$
where
$A=e^{-2iguFsu}-1$, $D=e^{-2iguFs}-1$,
$k_{\perp}^{2}=k_{1}^{2}+k_{2}^{2}$ and
$$\zeta=\frac{ch(gH_{3}s)-ch(gH_{3}(1-2u)s)}{sh(gH_{3}s)}.$$
Note  that $$\langle e^{-s\bar{P}^2}\rangle=
\frac{1}{(4\pi s)^{3/2}} \frac{gH_{3}s}{sh(gH_{3}s)}.$$
Finally, after  integrating by parts, we arrive at:
\begin{eqnarray*}
I_{ij}=\frac{g^{2}}{8\pi^{3/2}}\frac{T}{\sqrt{gH_{3}}}
\int\limits_0^1 d u
\int\limits_0^\infty \frac{d x}{sh(x)} \sqrt{x}~~ exp[-\Phi-x m]
M_{ij}(x,u).
\end{eqnarray*}
Here designations (\ref{five}) are used and
$x=gH_{3}s$. The matrix $M_{ij}$ is defined by
\begin{eqnarray*}
M_{ij}=
\Bigl\{2ch(2x)\Bigl
[(\rho\bar{k})_{i}~(\rho\bar{k})_{j}-\rho_{ij}(\bar{k}\rho\bar{k})+(\lambda\bar{k})_{i}(\lambda\bar{k})_{j}\Bigr]
+8(B\bar{k})_{i}(B\bar{k})_{j}\zeta sh(2x)
~~\nonumber\\
+4\Bigl[-(\Lambda(\sigma')\bar{k})_{i}(\Lambda(\sigma)\bar{k})_{i}-(\Lambda(-\sigma)\bar{k})_{i}(\Lambda(-\sigma')\bar{k})_{i}\nonumber\\
+\Lambda_{ij}(\sigma')(\bar{k}\Lambda(\sigma)\bar{k})+\Lambda_{ij}(-\sigma)(\bar{k}\Lambda(\sigma')\bar{k})\Bigr]\Bigr\},
\end{eqnarray*}
where $\sigma=2x(1-u)$, $\sigma'=2xu$, $\rho=(1-2u)R-\xi B^{2}$, $\lambda=\zeta B$, $\xi=\frac{sh(x(1-2u))}{sh(x)}.$
It can be easely verified that quantity $I_{ij}$ is manifestly transversal, $k_{i}I_{ij}=k_{j}I_{ij}=0$,
as it should be due to the gauge invariance.

Now we can apply described above procedure to evoluate the 
$\Pi_{ij}^{a=3,8}$. The result is given by:
\begin{eqnarray*}
\Pi_{ij}^{a=3}=I_{ij}+\frac{1}{6}\Pi_{ij}^{a=8},
\end{eqnarray*}

\begin{eqnarray*}
\Pi^{(i)}_{a=8}=\frac{3 g^{2}T}{16 \pi^{3/2}}
\int\limits_0^1 d u
\int\limits_0^\infty \frac{d x}{sh(x)} \sqrt{x}~
\Bigl\{\frac{1}{\sqrt{gH_{+}}}e^{-\Phi_{+}-xm_{+}}+\frac{1}{\sqrt{gH_{-}}}e^{-\Phi_{-}-xm_{-}}
\Bigr\}M_{ij}(x,u).
\end{eqnarray*}
It is conveniant to rewrite the operators $\Pi_{ij}^{a=3,8}$, using their eigenvectors, $b^{\rho}_{i}$, and eigenvalues, $\kappa_{a=3,8}^{\rho}$, as:
\begin{eqnarray*}
\Pi_{ij}^{a=3,8}=\sum_{\rho=1}^{3} \kappa_{a=3,8}^{\rho}
\frac{b^{\rho}_{i}b^{\rho}_{j}}{|b^{\rho}|^{2}},~~~~~~~~~~~~
\end{eqnarray*}
\begin{eqnarray*}
\Pi_{ij}^{a=3,8}b^{\rho}_{j}= \kappa_{3,8}^{\rho}
b^{\rho}_{i},~~~~~~~~~~~~
\end{eqnarray*}
where $b^{\rho=1}_{i}=(B\bar{k})_{i}$, $b^{\rho=2}_{i}=(R\bar{k})_{i}+\frac{k_{3}^{2}}{k_{\perp}^{2}}(B^{2}\bar{k})_{i}$ and $b^{\rho=3}_{i}=k_{i}$. The eigenvectors $b^{\rho}_{i}$ satisfy to the condition of completness:
\begin{eqnarray*}
\sum_{\rho=1}^{3} \frac{b^{\rho}_{i}b^{\rho}_{j}}{|b^{\rho}|^{2}}=\delta_{ij}.~~~~~~~~~~~~
\end{eqnarray*}
Hence, since $\kappa^{\rho=3}=0$ because of transversality of the $\Pi_{ij}^{a=3,8}$, we obtain  (\ref{six}).

\section{Appendix 2}
The functions $f^{(i)}(x,u)$ and $l^{(i)}_{\pm}(x,u)$ are:
\begin{eqnarray*}
f^{(1)}=4\Bigl[2ch(x)ch(x[1-2u])-\frac{1}{2}(1-2u)\xi ch(2x)\Bigr],~~~~~~~~~~~~
\end{eqnarray*}
\begin{eqnarray*}
f^{(2)}=4\Bigl[ch(2x)-2sh(2x)\zeta-\frac{1}{2}ch(2x)(\xi^{2}-\zeta^{2})\Bigr]-f^{(1)},~~~~~~~~~~~~
\end{eqnarray*}
\begin{eqnarray*}
f^{(3)}=4gA_3\Bigl[2sh(2x)-ch(2x)\zeta\Bigr],~~f^{(4)}=8(gA_3)^{2}ch(2x),
\end{eqnarray*}
\begin{eqnarray*}
l^{(i=1,2)}_{\pm}=f^{(i=1,2)},~~~l^{(3)}_{\pm}=\frac{A_{\pm}}{A_3}f^{(3)},~~~l^{(4)}_{\pm}=\frac{(A_{\pm})^{2}}{(A_3)^{2}}f^{(4)}
\end{eqnarray*}
and
$$\xi=\frac{sh(x(1-2u))}{sh(x)},~~~\zeta=\frac{ch(x)-ch(x(1-2u))}{sh(x)}.$$

\newpage
\begin{figure}\centering
\includegraphics[scale=0.4]{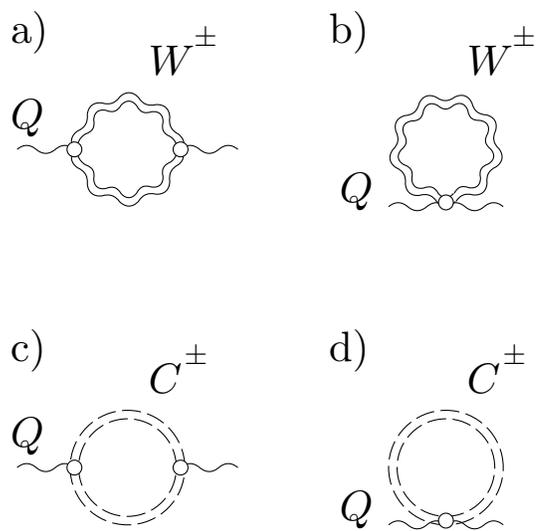}
\caption{Polarization operator of neutral gluons in the one-loop approximation.} \label{Fig1}
\end{figure}

\end{document}